\documentclass[graybox]{svmult}
\pdfoutput=1

\usepackage{mathptmx}       
\usepackage{helvet}         
\usepackage{courier}        
\usepackage{type1cm}        
\usepackage{makeidx}         
\usepackage{graphicx}        
\usepackage{multicol}        
\usepackage[bottom]{footmisc}
\usepackage{algorithm}
\usepackage{algorithmic}
\usepackage{longtable}
\usepackage{enumitem}
\usepackage{supertabular,booktabs}
\usepackage{tikz}
\usepackage{caption,subcaption}  
\usepackage{subcaption}
\usepackage{caption}
\usepackage{comment}
\usepackage{csquotes}
\usepackage{tablefootnote}
\usepackage{amsmath}
\usepackage{amssymb}
\usepackage{mathtools }
\usepackage[numbers]{natbib}
\usepackage{tikz}
\usetikzlibrary{arrows,automata, positioning, patterns,arrows.meta}
\usepackage{pgfplots}
\usetikzlibrary{datavisualization}
\usepgfplotslibrary{groupplots}

\makeindex             
\newcommand{\naturalnumber}{\mathbb{N}} %
\newcommand{\argument}{\Phi}
\newcommand{\arguments}{\textit{Args}}

\newcommand{\feedback}{\textit{f}\,}

\newcommand{\component}{\varphi}

\newcommand{\effectiveness}{\textit{e}}

\newcommand{\clang}{L_c}

\newcommand{\tlang}{L_t}

\usepackage{array}
\newcommand{\lvlfocus}{focus}
\newcommand{\userfocus}{\mathcal{F}}
\newcommand{\pair}{\ddot{\textit{P}}}
\newcommand{\single}{\dot{\textit{P}}}
\newcommand{\all}{\textit{P}}

\newcommand{\rue}{RUE}
\newcolumntype{P}[1]{>{\arraybackslash}p{#1}}
\newcolumntype{M}[1]{>{\arraybackslash}m{#1}}
\newcolumntype{?}{!{\vrule width 1.5pt}}
\DeclareMathOperator*{\argmax}{arg\,max}

\definecolor{graph_red}{rgb}{1.0,0.35,0.35}
\definecolor{graph_green}{rgb}{0.25,1.0,0.25}

\begin{document}

\title*{Fostering User Engagement in the Critical Reflection of Arguments}
\author{Klaus Weber, Annalena Aicher, Wolfgang Minker, Stefan Ultes and Elisabeth Andr\'e}
\authorrunning{Weber et al.} 
\institute{Klaus Weber \at University of Augsburg, Germany, \email{klaus.weber@uni-a.de}
\and Annalena Aicher \at University of Ulm \email{annalena.aicher@uni-ulm.de}}
%
%
\maketitle

\abstract*{A natural way to resolve different points of view and form opinions is through exchanging arguments and knowledge. Facing the vast amount of available information on the internet, people tend to focus on information consistent with their beliefs. Especially when the issue is controversial, information is often selected that does not challenge one’s beliefs. To support a fair and unbiased opinion-building process, we propose a chatbot system that engages in a deliberative dialogue with a human. In contrast to persuasive systems, the envisioned chatbot aims to provide a diverse and representative overview - embedded in a conversation with the user. 
To account for a reflective and unbiased exploration of the topic, we enable the system to intervene if the user is too focused on their pre-existing opinion. Therefore we propose a model to estimate the users’ reflective engagement (RUE), defined as their critical thinking and open-mindedness.
We report on a user study with 58 participants to test our model and the effect of the intervention mechanism, discuss the implications of the results, and present perspectives for future work. The results show a significant effect on both user reflection and total user focus, proving our proposed approach's validity.}

\abstract{A natural way to resolve different points of view and form opinions is through exchanging arguments and knowledge. Facing the vast amount of available information on the internet, people tend to focus on information consistent with their beliefs. Especially when the issue is controversial, information is often selected that does not challenge one’s beliefs. To support a fair and unbiased opinion-building process, we propose a chatbot system that engages in a deliberative dialogue with a human. In contrast to persuasive systems, the envisioned chatbot aims to provide a diverse and representative overview - embedded in a conversation with the user. 
To account for a reflective and unbiased exploration of the topic, we enable the system to intervene if the user is too focused on their pre-existing opinion. Therefore we propose a model to estimate the users’ reflective engagement (RUE), defined as their critical thinking and open-mindedness.
We report on a user study with 58 participants to test our model and the effect of the intervention mechanism, discuss the implications of the results, and present perspectives for future work. The results show a significant effect on both user reflection and total user focus, proving our proposed approach's validity.}

\section{Introduction}

Discussing arguments and resolving different viewpoints is essential for humans to build and form an opinion. With the rapid spreading of social media platforms and various other possibilities the internet provides to exchange arguments, a shift from face-to-face to online discussion is perceivable. 
The overwhelming amount of information makes it difficult to sort through and establish a well-founded opinion. Furthermore, the information presented by search engines, social media, etc.\ has often been filtered and reduced by algorithms that consider previous user queries. As a result, it is even more likely that people tend to focus on sources that repeat and reinforce a pre-existing opinion. 
Exploring the opposing view requires more effort; thus, people are more likely to be one-sided on their view~\cite{gelter_why_2003}, which contradicts \emph{reflective thinking}. Reflective thinking needs an ongoing reevaluation of one's beliefs, assumptions, and hypotheses~\cite{king1994developing}.
According to~\cite{paul_critical_1990}, if people think critically in a \emph{weak sense} only, they tend to use skills to defend what is already in their thoughts. In other words: \emph{weak sense} implies reflecting about positions that are different from the own ones~\cite{mason2007critical}, but tending to defend the own view without reflection~\cite{paul_critical_1990}, and \emph{strong sense}  means to be able to reflect one's own opinions as well. The required energy and effort~\cite{gelter_why_2003} for this reflection is often absent due to a lack of people's \emph{need for cognition}~\cite{maloney2020higher}. Due to the users' tendency to defend their view~\cite{paul_critical_1990}, a debating system that confronts them with an opposing stance might not lead to critical reflection but rather the opposite. Thus, a respective argumentation system should enable users to freely explore their own and opposing views without being directly confronted in a debate. 

In order to find an objective measure for the user's reflective engagement (RUE\footnote{Within this paper, reflective engagement (RE) and user's reflective engagement (RUE) describe the same phenomenon. As RUE refers to an individual user, we chose this term for our user model.}) we consider the explicitly expressed user stance on the topic in relation to the user's focus. Consequently, our proposed model yields a low RE if the users focus on arguments that align with their stance. This model is included in the first working prototype for a dialog system that intervenes during the conversation to motivate and encourage the user to overcome a less reflective argument exploration. This means that whenever the system notices the user to be focused too firmly on their pre-existing opinion, the system suggests an argument contradicting their opinion.
To evaluate the feasibility and validity of our proposed intervention approach, we conducted an online user study with 58 participants who interacted with our argumentative system. The results show that the intervention significantly affects both the user reflection and total user focus, proving our model's validity.
The remainder of the paper is as follows: Section~\ref{sec:related_work} gives an overview of the related work w.r.t. argumentation systems and reflective engagement. The formal argument structure, dialogue framework, and its modules are described in Section~\ref{sec:framework}. In Section~\ref{sec:re_model}, we discuss the derivation of the user's reflective engagement. We present the results of the user study in Section~\ref{sec:evaluation}. 

\section{Related Work}
\label{sec:related_work}
In the following, we cover the related work of 1) argumentative dialog systems and 2) reflective engagement. 

\subsection{Argumentative Dialog Systems}
Argumentative dialog systems, conversational agents (CA), and Chatbots aim to interact with users through natural language by exchanging arguments. 
Most approaches to human-machine argumentation are embedded in a competitive setting~\cite{slonim21-AAD,rosenfeld16-SAA}. They utilize different models to structure the interaction. Arguing chatbots, like Debbie, use a similarity algorithm to retrieve counterarguments~\cite{rakshit2017-DTD}, and Dave uses retrieval- and generative-based models~\cite{Le18-DTD} interacted via text with the user. A menu-based framework that incorporates the beliefs and concerns of the opponent was presented by~\cite{hadoux2021-SAD}. In the same line,~\cite{Chalaguine20-APC} used a previously crowd-sourced argument graph and considered the user's concerns to persuade them. A persuasive prototype chatbot is introduced by~\cite{Chalaguine21APC} to convince users to vaccinate against COVID-19 using computational models of argument.
In contrast to all ADS mentioned above, ~\cite{aicher2021-OBB} introduces an argumentative dialogue system that provides arguments upon users' request. They pursue a cooperative exchange of interesting arguments without trying to persuade or win a debate against the user. We adopt this cooperative approach, as according to~\cite{hart2009feeling}, a mere confrontation with opposing arguments leads to cognitive dissonance, which can have a negative effect~\cite{harmon-jonesCDENA2000} e.g. in the form of evoking a defense motivation in individuals. Therefore, a confrontation in a competitive scenario is more likely to lead to rejection rather than an open-minded attitude. Furthermore, it is crucial to maintain the user's trust~\cite{kraus2021role} and enable the user to decide whether to accept or decline an argument suggestion. 

\subsection{Reflective Engagement}
Most of the literature on reflective engagement (RE) refers to learners' continual and active participation in their problem inquiry with a continuous and critical judgment of the inquiry process and inquiry outcomes for possible improvement.~\cite{farr2012-SEI} explored reflective markers for examining the practice of reflection for educational purposes in online and face-to-face discussions (interactive) as well as blogs and essays (non-interactive). They identified verbs such as \emph{think} and \emph{know}, adverbs such as \emph{very} and \emph{bit}, adjectives such as \emph{good} and \emph{sure} and finally, nouns such as \emph{problem}, and \emph{fact} as most-used ones.  
According to~\cite{lyons2006-REA}, who reflects on teaching practices, RE involves deliberately interrupting one's own teaching practice to question it systematically. Therefore, the teacher's conscious awareness of their practices and students should be redirected to adapt and change their practices accordingly. ~\cite{rodman2010-FTT} defined a continuum of reflective engagement from a teacher-centered to a student-centered focus. She states that reflective engagement is crucial to a teacher's understanding of the nature of teaching and learning. Thus, most of the literature~\cite{kong2015-AEO,silpasuwanchai2016-DAC} investigates the RE in the context of teaching-learning processes, whereas we focus on the exploration of diverging points of view in argumentative scenarios.
Contrary to the approach of~\cite{farr2012-SEI}, who measure reflective thinking by means of identified markers or~\cite{aicher_determination_2021}, who propose a balanced determination of the user's reflective engagement, we take the user's stance as well as the explored side (polarity) of the argumentation into account to calculate the reflective engagement, in line with~\cite{paul1981teaching,mason2007critical}. 
\cite{aicher_determination_2021} proposed a first derivation of the reflective user engagement score. They consider a user's reflective engagement as high if the user explores both sides of the argumentation. Whereas their approach covers the balanced exploring behavior of the user, we introduce a positive reward if the user scrutinizes the view differing from their own (current) opinion, which means a more vital critical reflection by the user~\cite{paul_critical_1990,gelter_why_2003}. Moreover, their RUE score does not consider which stance the user focuses on more in case of an imbalanced exploration. Hence, we extend their approach by estimating the user's stance and derive an RUE model using the negative correlation between the user's stance and focus.

\section{Prototype \& Architecture}
\label{sec:framework}
In the following, we give a short overview of the dialog framework and the intervention prototype setup of our argumentative dialog system. We employ a modified version of the dialogue framework introduced by~\cite{aicher2021-OBB}.
\subsection{Argument Structure}
\begin{figure}[htp]
	\centering
	\includegraphics[width=0.65
	\columnwidth]{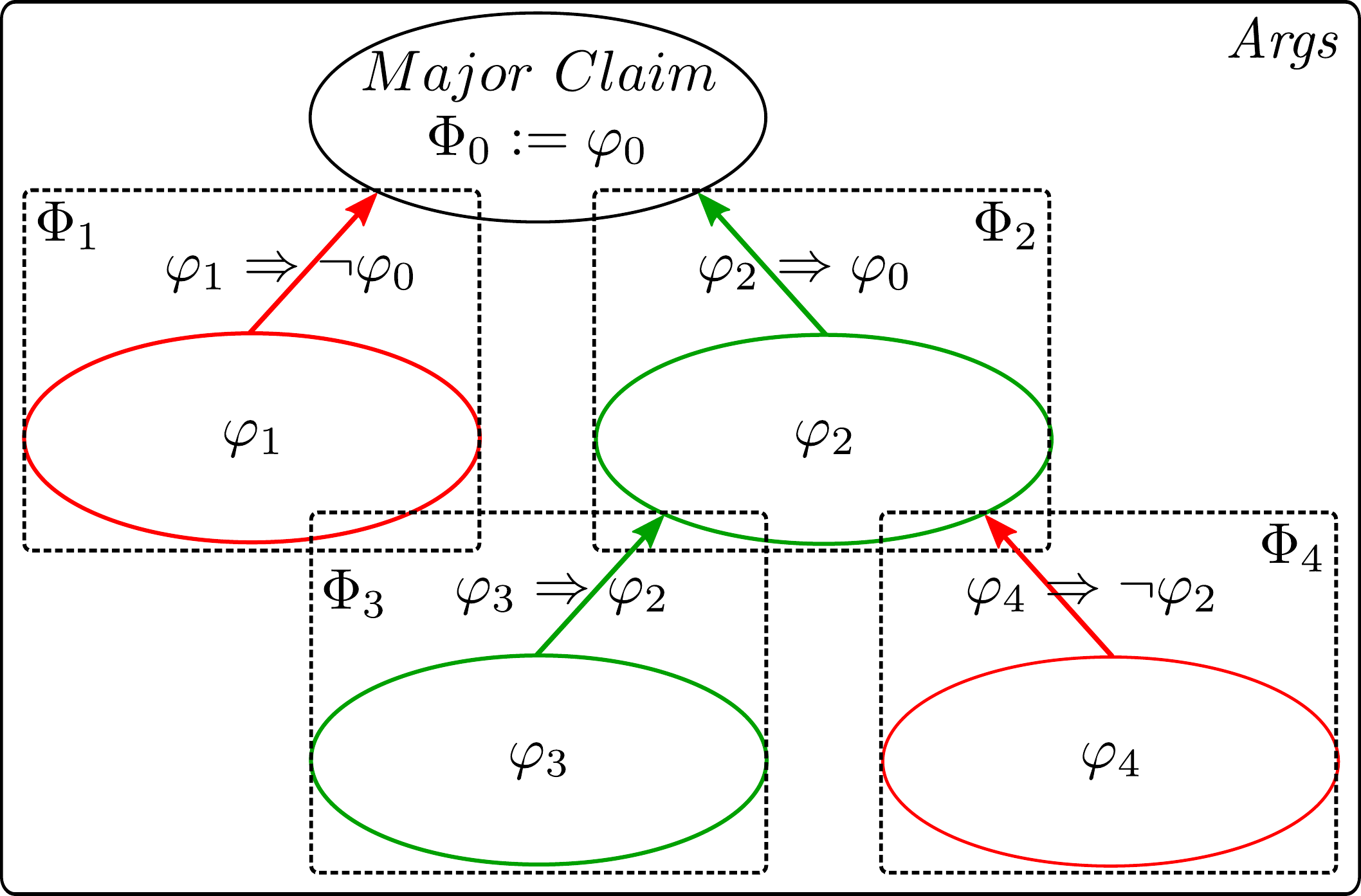}
	\caption{Argument tree structure consisting of components $\component_i$ and $\component_j$ as well as logical statements between them.}
	\label{fig:argument_strucuture}
\end{figure}
The dialog framework consists of components $\tlang$, and a communication language $\clang$. Each component $\component_i \in \tlang$ has a unique textual representation and a relation (\emph{support}, \emph{attack}) to exactly one other component (logical inference) to define arguments $\argument_i \in \arguments$, such that  
it is $\argument_i = \left(\component_i \Rightarrow \neg\component_j\right)$ an \emph{attacking} argument or $\argument_i = \left(\component_i \Rightarrow \component_j\right)$ a \emph{supporting} argument resulting in a bipolar argument tree structure (Fig. \ref{fig:argument_strucuture}). Throughout this work, we use the \emph{idebate} dataset \emph{Marriage is an outdated institution}\footnote{\url{https://idebate.org/debatabase}} consisting of 72 arguments following the presented structure. 
The root argument (\emph{Major Claim}) is defined as $\argument_0 := \component_0$.  Every argument $\argument_i \in \arguments$ has a stance $\in \{+, -\}$ towards $\component_0$ defined by the component relation and the respective position in the argument graph. 
The communication language $\clang$ includes the speech acts available to the user and system (see Tab.~\ref{tab:speechacts}). 

\begin{table}[h!]
	\centering
	\begin{tabular}{P{2.5cm} |P{6.5cm}}
		\toprule
		{Speech Act} &  {Description}\\ \midrule\midrule
		\multicolumn{2}{c}{ System moves}\\ \midrule
		\emph{$argue(\component_i \Rightarrow \component_j$)} & {Present argument $\component_i  \Rightarrow \component_j$}\\
		\emph{$jump\_{to}(\component_i)$} & {Jump to argument $\argument_i = \component_i \Rightarrow *$\tablefootnote{$* \in \{\component, \neg \component\}$}}\\
		\emph{$intervene$} & {Intervene last user choice}\\
		\midrule
		\multicolumn{2}{c}{ User moves}\\ \midrule
		\emph{$why_{pro}(\component_i$)} & {Ask for a supporting component $\component_j$ with  $\component_j \Rightarrow \component_i$}\\
		\emph{$why_{con}(\component_i$)} & {Ask for an attacking component $\component_j$ with  $\component_j \Rightarrow \neg\component_i$}\\
		\emph{$level_{up}$} &  {Move level up}\\
		\emph{$agree(\component_i)$} & {Feedback to agree with a statement $\component_i$} \\
		\emph{$disagree(\component_i)$} & {Feedback to disagree with a statement $\component_i$} \\
		\emph{$confirm/reject$} & {Confirm/Reject intervention} \\ 
		\bottomrule
	\end{tabular}
	\caption{\label{tbl-comlang} Communication language $\clang$ of the herein implemented dialogue framework consisting of nine speech acts.}
	\label{tab:speechacts}
\end{table}


\subsection{System Architecture}
\begin{figure*}[h!]
	\centering
	\includegraphics[width=1.0
	\columnwidth]{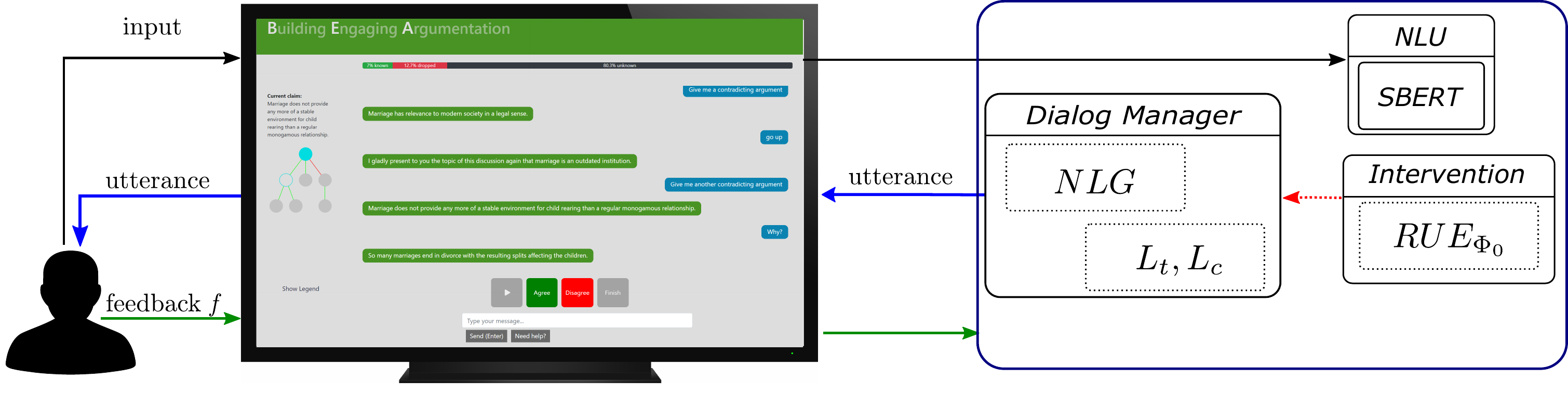}
	\caption{System overview for a user interacting with the system via chat input/output.}
	\label{fig:system}
\end{figure*}

Figure~\ref{fig:system} sketches the architecture of our system. It consists of 1) an NLU (Natural Language Understanding), 2) a dialog manager, 3) an intervention module, and 4) an NLG (Natural Language Generation).

\subsubsection{Natural Language Understanding} 
The system has a chat-based input field where users can freely type in their requests to allow for a natural conversation. 
An integrated natural language understanding framework (NLU)~\cite{abro21-NLU} is required to map this input to the available speech acts. It uses an intent classifier model consisting of two main components: a BERT Transformer Encoder and a bidirectional LSTM classifier~\cite{abro_natural_2021}. Furthermore, to identify the arguments a user refers to, a similarity model is applied, which is based on Sentence-BERT (SBERT)~\cite{reimers2019sentence}. 

\subsubsection{Dialog Manager} 
The dialog manager consists of the argument tree (components $\tlang$) and the communication language $\clang$. It manages the dialog flow between the system and the user and ensures logical consistency throughout the dialog. 
In addition, the dialog manager stores the current dialog state, i.e., which arguments have been presented, the current position within the argument graph, and allowed speech acts. For instance, if an argument $\argument_i$ is a leaf node, $why_{pro}(\component_i)$ is not allowed. If the user requests a new argument ($why_{x}$), the system selects a random argument from all arguments fitting the requested relation $x \in\{\emph{pro}, \emph{con}\}$.



\subsubsection{Natural Language Generation} 
The system provides two output modalities: a \emph{textual} or uttered \emph{speech response}. The NLG is based on the original textual representation of the argument components $\component_i \in \tlang$.
The annotated sentences were manually modified in terms of grammatical syntax to form stand-alone utterances, which serve as a template for the respective system response. A list of natural language representations for each speech act is also defined. During the response generation, the explicit formulation is chosen from this list randomly. To indicate that the system does not represent the arguments but merely reflects them,  formulations, such as \textit{This claim is supported by the argument that...} are chosen. For better comprehensibility, the system describes its actions, e.g., by stating~\textit{Let us return to the previous argument, that...} when navigating to the parent node.

\subsubsection{Intervention module}
The intervention keeps track of the user's reflective engagement (RUE) ($\rue_0$, see Sec.~\ref{sec:focus_rue} for calculations) and intervenes if necessary. I.e., it suggests an argument with a different stance than requested by the user. Let $l^+, l^- \in \tlang$ be the set of all valid\footnote{A valid pro (+) and con (-) argument component $\component_i$ is an unheard component that links to any component $\component_j$ that has already been presented to the user.} argument components that the system can present. For every user request, the system simulates the $\rue_0$ for all valid arguments with opposite stances and returns the component that maximizes it. The intervention takes place if and only if the simulated RUE is greater than the maximum possible RUE for all available arguments that match the user request, i.e., if the user requests a supporting argument, the intervention module checks if 
\begin{equation}
	\max_{\component_i \in l^-}(sim\_rue(\component_i)) > \max_{\component_i \in l^+}(sim\_rue(\component_i))
\end{equation}
and, if so,  returns $\component_i = \argmax_{\component_i \in l^-}(sim\_rue(\component_i))$, and suggests it to the user, who can accept or deny it. In case of denial, the system proceeds with the initial user request. 

\subsection{User Interface}

\begin{table}[t!]
	\begin{center}
		\begin{tabular}{P{1cm}| p{8cm}| P{2.3cm}}
			\toprule
			{Actor} &  {Utterance} &  {Speech Acts}\\ 
			\midrule
			{system}  & \emph{Let's discuss that marriage is an outdated institution.} & claim($\component_0$) \\
			\midrule
			{User}  & Tell me more & {$why_{pro}(\component_0$)} \\
			\midrule
			{system}  & \emph{Marriage does not provide any more of a stable environment for child rearing than a regular monogamous relationship.} & {$argue(\component_1\Rightarrow~\component_0$)}\\
			\midrule
			User  & I agree & $agree(\component_1$)\\
			\midrule
			{system}  & \emph{Alright, I noted that you agree with the presented statement. Let us return to the topic of this discussion that marriage is an outdated institution} & - \\
			\midrule
			{User}  & Give me another supporting argument &  {$why_{pro}(\component_0$)} \\
			\midrule
			{system}  & \emph{I think we should look at the opposite point of view. Alright?} &   {$intervene$}\\
			\midrule
			{User}  & {Yes} &  {$confirm$}  \\
			\midrule
			{system}  & \emph{Marriage removes the transient and casual aspects of a monogamous relationship, thus giving a child a far more stable environment.} &  {$argue(\component_2\Rightarrow~\neg\component_0$)}\\
			\bottomrule
		\end{tabular}
	\end{center}
	\caption{\label{tab:dialogue_ex} Artificial dialogue between the system and the user about the topic \emph{\enquote{Marriage is an outdated institution}}.}
\end{table}

\begin{figure}[htp]
	\centering
	\includegraphics[width=0.75
	\columnwidth]{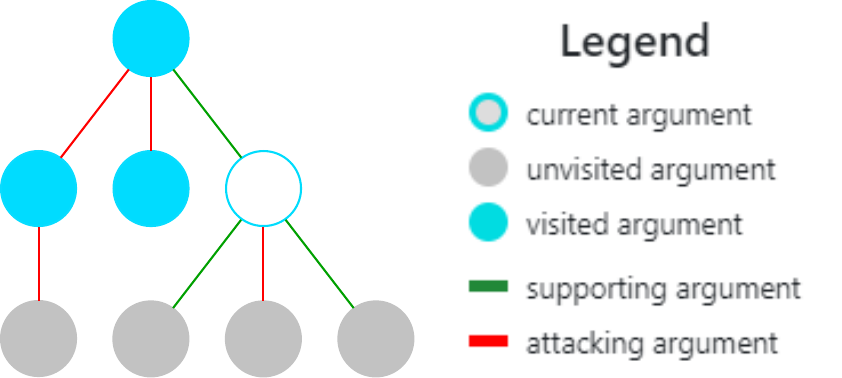}
	\caption{Sketch of the subgraph displayed to the user on the left side of the browser window.}
	\label{fig:nodes}
\end{figure}

The user interface consists of four components: 1) the system output, 2) the user input, 3) the user feedback buttons, and 4) the graphically displayed information about the argument structure.
The system's \emph{textual} response is displayed in a classic chat design. The system provides a \emph{text} (chat) input where users can freely type in their requests. To allow for feedback on whether users agree with the current argument or not, there are two buttons (\emph{Agree} and \emph{Disagree}) that the user can use at any time during interaction. Without feedback, the system considers a neutral user stance. 
The allowed user speech acts strongly depend on the user's position in the argument graph. Moreover, our model assumes that unheard arguments were omitted deliberately and not accidentally. Therefore, the current subgraph is displayed on the left side of the browser window, as can be seen in Fig.~\ref{fig:nodes}. 

The subgraph is a visual representation of all arguments of the respective claim\footnote{Here, a claim $\argument_j$ directly \emph{attacks} or \emph{supports} the Major claim $\argument_0$.} 
The subgraph is generated automatically using a visualization library processing the adjacent matrix representation of the argument graph. The user's current position is denoted by a \emph{blue outlined node}, the already discussed arguments are shown in \emph{blue}, and unheard arguments in \emph{gray}. The edges between the nodes show a \emph{supporting} relation in \emph{green} and an \emph{attacking} relation in \emph{red} color. Users have the free choice of whether or not they want to ask for a pro or con argument or how they want to navigate through the argument tree. It is noteworthy that intervention only takes place if and only if the user input is mapped to either \emph{$why_{pro}$} or \emph{$why_{con}$}. Table~\ref{tab:dialogue_ex} sketches a sample dialogue between the user and system.

\section{Reflective Engagement Model}\label{sec:re_model}
Within this section, we describe the derivation of the underlying theoretical model in detail.
As aforementioned, we extend the approach of \cite{aicher_determination_2021} by an estimation model of the user's stance (denoted as $\effectiveness_0 \in [0,1])$ and focus and derive an RUE model using the negative correlation between the user's stance and focus. 

\subsection{User Stance Estimation}
To obtain an accurate estimation of the user's stance, we employ a model proposed by~\cite{weber2020predicting}. During the interaction with the system, users can give feedback on whether or not they agree or disagree with any argument $\argument_i \in \arguments$. Considering the hierarchical structure of arguments, the system uses this feedback to compute the user's stance $\effectiveness_{\argument_j} \in [0,1]$ of any argument $\argument_j \in \arguments$ considering the feedback for this respective argument and the feedback for all arguments in the subtree with $\argument_j$ as root. Thereby, the overall stance $\effectiveness_{\argument_0}$ of the discussed topic is computed.

\subsection{Focus and Reflective Engagement Score}
\label{sec:focus_rue}
To derive our reflective engagement score, we first define the user's focus and derive the reflective engagement score by considering the predicted user's stance and its correlation to the user's focus. 
A set of arguments with the same target argument $\argument_i$ is denoted as $\all_{\argument_i}$. 
If it is in favor of stance +, it is denoted as $\all_{\argument_i}^{+}$ and $\all_{\argument_i}^{-}$ elsewise. 
The set of all visited arguments $\argument_j$ with target argument $\argument_i$ is denoted as $\all_{\argument_i,v}$.
We define the user's focus for argument $\argument_i$ based on visited pro and con arguments as:
\begin{equation}
	\lvlfocus_{\argument_i} =  \frac{\left|\all_{{\argument_i},v}^+\right| - \left|\all_{\argument_i,v}^-\right|}{\left|\all_{\argument_i,v} \right|} \in [-1,1]
\end{equation}
It is easy to verify that the more arguments of a certain stance are selected by the user, the more the focus shifts in the direction of the respective stance. 

As we use hierarchical argumentation structures, we adopt the approach of \cite{aicher_determination_2021} to take the former into account by introducing a weight respective to the "position" of the target argument on the one hand and a weight for the size of the underlying structure of the target argument on the other. Thus, a hierarchical weight $\omega_{d,i}$ incorporates the different levels of argument depth into the reflective engagement. This ensures that exploring lower levels will be assigned larger weight values than near the target argument~\cite{aicher_determination_2021}. This choice is based on the fact that arguments are mapped out due to the hierarchical argument structure, i.e., every node within the hierarchy contains additional evidence for the overall proposition that directly attacks/supports the Major Claim.
\begin{equation}\label{eq:omegad}
	\omega_{d_{(k-i)}} = \frac{d_{k-i}}{\sum_{m=1}^{l_{\text{max}}-i} m}, 
\end{equation}
where $d_{(k-i)}$ denotes the depth of the level $i$ with respect to the level of target argument at level $k$.
To avoid an over-representation of levels with only a few arguments while levels with many arguments will be under-represented, the second weight $\omega_{n}$ is introduced. It takes into account the different sizes of levels beneath the target argument $\argument_i$. Thus, we relate the number of descendants of the respective level $k$ to all descendants such that  
\begin{equation}\label{eq:omegad}
	\omega_{n, \argument_i} = \frac{\big|\all_{\argument_i}\big|} {\sum_{\argument_m\in des(\argument_i)} \big|\all_{\argument_m}\big|}, 
\end{equation}
where $des: \arguments~\rightarrow~P(\arguments)$ is the set of all arguments in the subgraph excluding leaf nodes with a given argument $\argument_i$ as root. Leaf nodes are excluded to avoid zero division. In addition, since leaf nodes have no descendant, thus,  $\omega_{n, \argument_i}$ is undefined for nodes. Further, let $lvl: \arguments~\rightarrow~\naturalnumber$ be the respective level of the argument in the argumentation structure. 
The total normalized user focus $\userfocus_{\argument_i} \in [-1,1]$ is then defined incorporating the defined weights as follows.
\begin{equation}
	\userfocus_{\argument_i} := \frac{\sum_{\argument_k\in des(\argument_i)} focus_{\argument_k} \cdot W_{\argument_k}}{\sum_{\argument_k\in des(\argument_i)} W_{\argument_k}}
\end{equation}
with $W_{\argument_k} = \omega_{d_{(lvl(\argument_k)+1-lvl(\argument_i))}}\omega_{n,\argument_k}$.
The user's reflective engagement considers the weighted user's focus by making use of the negative correlation of stance and focus. 
This approach ensures that if the user stance is positive ($+$), the system intervenes to suggest the user choose more con arguments and vice versa. 

\begin{equation}
	\rue_{\argument_i} = 1 - \left|\effectiveness_{\argument_i} - \left(1 - \frac{\userfocus_{\argument_i} + 1}{2}\right)\right|
\end{equation}
After inverting the normalized focus, the difference of user stance and focus is taken such that the closer the absolute value of the difference is to 0.0, the higher $\rue_{\argument_i}$.
In other words: 1) The closer the current user's stance is to 0.5, the more likely both argument stances are allowed (i.e., $\userfocus_{\argument_i} \approx 0.0$). The closer the current user's stance is to 0 or 1, the more single arguments contrary to the user's stance are necessary to keep the reflective user engagement up. This inevitably gives the system a well-fitting score to keep track of the user's reflection and adopt a strategy that increases reflective user engagement. 

\begin{figure*}[htp]
	\centering
	\includegraphics[width=1.0
	\textwidth]{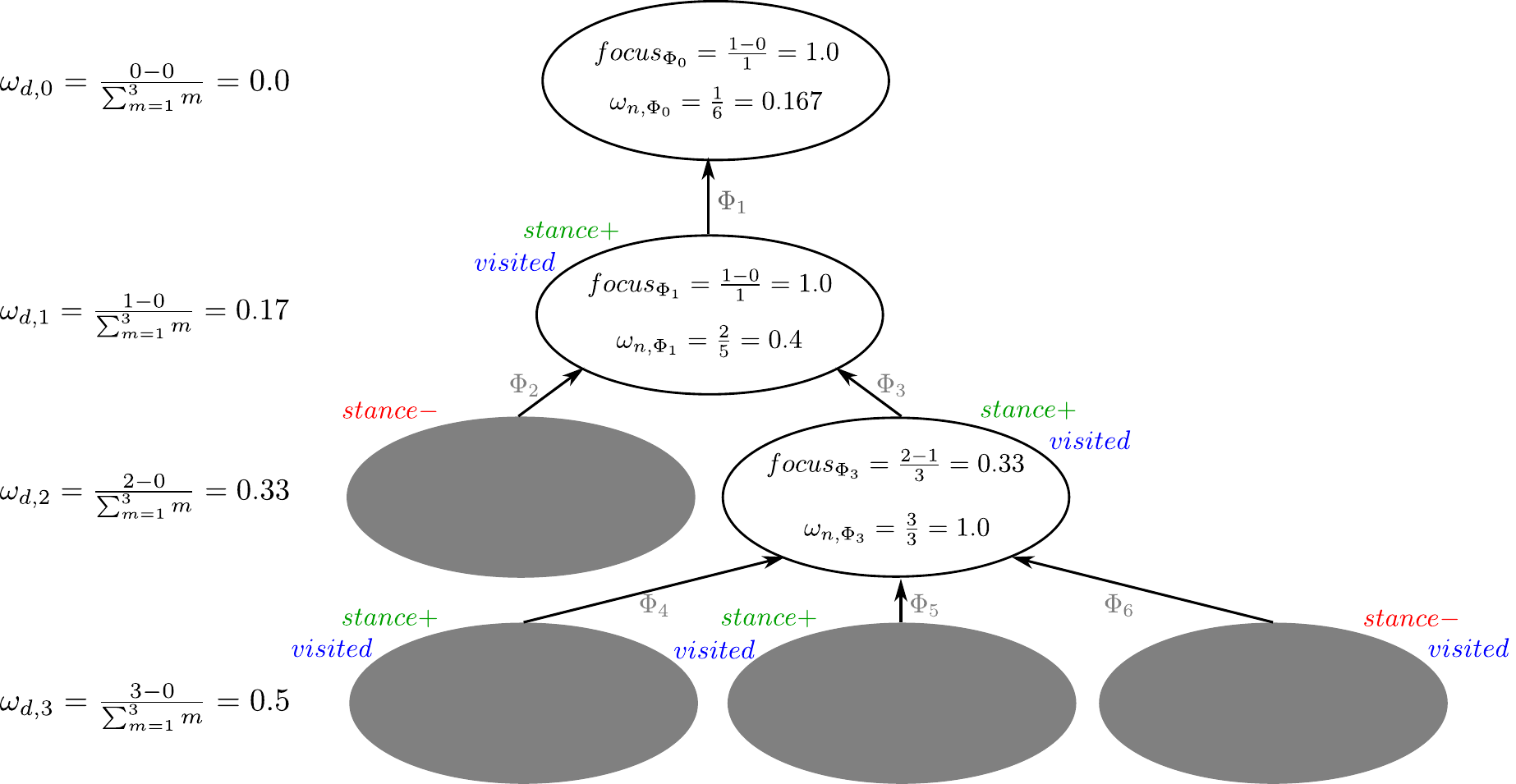}
	\caption{Example: Several arguments either with stance \emph{+} or \emph{-} and information whether or not the argument is visited by the user and the computed $\omega_{d_{(k-i)}}, \omega_{n, \argument_i}$, and $focus_{\argument_i}$ values. Leaf nodes are shown in gray.}
	\label{fig:rue_example}
\end{figure*}

Fig.~\ref{fig:rue_example} sketches an example. It depicts several arguments either with stance \emph{+} or \emph{-} and information on whether or not the user visited the argument. Following the formulae above, we first compute the hierarchical weights $\omega_{d_{(k-i)}}$ for every level $k \in [0,3]$ and the descendant weights $\omega_{n, \argument_i}$ and $focus_{\argument_i}$ for every argument $\argument_i \in \arguments$ with $i \in [0,1,3]$. Arguments $\argument_2, \argument_4, \argument_5, \argument_6$ are leaf nodes; thus, no descendant weights can be computed for them. Using these information, we compute the user focus $\userfocus_{\argument_0} = \frac{ 1 * 0.167 * 0.17 + 1 * 0.4 * 0.33 + 0.33 * 1 * 0.5}{0.167 * 0.17 + 0.4 * 0.33 + 1 * 0.5} = 0.49$ and with that the reflective user engagement $RUE_{\argument_0} = 1 - \left|0.0 - \left(1 - \frac{0.74 + 1}{2}\right)\right| = 0.745$ assuming a negative user stance of $\effectiveness_{\argument_0} = 0.0$. As the user selected mostly arguments with stance \emph{+} but also an argument with stance \emph{-}, the RUE score is 0.745. Without the negative argument, it is easy to verify that the RUE would be 1.0 in this example. 

\section{User Study}
\label{sec:evaluation}
We conducted a user study with 58 participants (aged 18-63, 50\% m/f) divided into two groups (an experimental group with intervention and a control group without intervention) to show the validity of the proposed user reflection score and the intervention mechanism. The study aimed to analyze the following questions: 
\begin{enumerate}
	\item Does the intervention affect user reflection?
	\item Does the intervention affect explored opposing arguments?
\end{enumerate}
More precisely, we defined the following hypotheses to be examined in the study: 
\begin{enumerate}[label=H\arabic*]
	\item \label{h:1} Participants in the experimental group (w/ intervention) were more reflective than in the control group.
	\item \label{h:2} The intervention mechanism and the proposed model lead to increased engagement with the arguments opposing the user's stance.
\end{enumerate}

\subsection{Participants, Apparatus, and Procedure}
The study was conducted online via the crowdsourcing platform "Crowdee\footnote{\url{https://www.crowdee.com/}}" with participants from the UK, US, and Australia (English native speakers to avoid language barrier effects). 
The study setup used the chat-based output modality. After an introduction to the system (short text and description of how to interact with the system), the users were advised to explore enough arguments to build a well-founded opinion on the topic \emph{Marriage is an outdated institution}. The participants were not told anything about the underlying reflection model but only to select at least ten arguments. In addition, they were asked to rate their opinion on the topic on a 5-point Likert scale, which normalized in $[0,1]$ displayed the initial user stance $\effectiveness_0$. 
During the study, we collected the following data anonymously based on privacy rights laws: 
\begin{enumerate}
	\item Measured user reflection score $\rue_0$  (Fig.~\ref{fig:data_a}).
	\item User stance $\effectiveness_0$ (Fig.~\ref{fig:data_c})
	\item Set of visited arguments $\all_{v}^+$ and $\all_{v}^-$  (Fig.~\ref{fig:data_d}).
\end{enumerate}
\begin{figure}[h!]
	\begin{subfigure}[b]{1.0\columnwidth}\centering
		\subcaption[]{User reflection score $\rue_{\argument_0}$}
		\begin{tikzpicture}[trim axis left,scale=1.0]
			\begin{axis}[
				height=4cm,
				width=9.6cm,
				grid=major,
				xmin=0,
				xmax=58,
				xlabel={Participant},
				ylabel={$\rue_{\argument_0}$},
				ytick distance=0.3,
				y label style={at={(axis description cs:0.08,.5)},anchor=south, font={\scriptsize}},
				x label style={at={(axis description cs:0.5,-0.15)},anchor=south,font={\scriptsize}},
				]
				\addplot[style={mark=x}, skip coords between index={30}{58}] table [x=id, y=rue, col sep=comma] {data.csv};
				\addplot[style={mark=o}, skip coords between index={0}{30}] table [x=id, y=rue, col sep=comma] {data.csv};
			\end{axis}
		\end{tikzpicture}
		\label{fig:data_a}
	\end{subfigure}%
	
	
	\begin{subfigure}[b]{1.0\columnwidth}\centering
		\subcaption[]{User stance $\effectiveness_{\argument_0}$}
		\begin{tikzpicture}[trim axis left,scale=1.0]
			\begin{axis}[
				height=4cm,
				width=9.6cm,
				grid=major,
				xmin=0,
				xmax=58,
				xlabel={Participant},
				ylabel={$\effectiveness_{\argument_0}$},
				ytick distance=0.3,
				y label style={at={(axis description cs:0.08,.5)},anchor=south, font={\scriptsize}},
				x label style={at={(axis description cs:0.5,-0.15)},anchor=south,font={\scriptsize}},
				]
				\addplot[style={mark=x},  skip coords between index={30}{58}] table [x=id, y=e_0, col sep=comma] {data.csv};
				\addplot[style={mark=o}, skip coords between index={0}{30}] table [x=id, y=e_0, col sep=comma] {data.csv};
			\end{axis}
		\end{tikzpicture}
		\label{fig:data_c}
	\end{subfigure}
	
	\begin{subfigure}[b]{1.0\columnwidth}\centering
		\subcaption[]{Ratio $\all_{v}^+:\all_{v}^-$}
		\begin{tikzpicture}[trim axis left,scale=1.0]
			\begin{axis}[
				height=4cm,
				width=9.6cm,
				grid=major,
				xmin=0,
				xmax=58,
				xlabel={Participant},
				ylabel={ratio in $\%$},
				ytick distance=0.3,
				y label style={at={(axis description cs:0.08,.5)},anchor=south, font={\scriptsize}},
				x label style={at={(axis description cs:0.5,-0.15)},anchor=south,font={\scriptsize}},
				]
				\addplot[style={mark=x}, skip coords between index={30}{58}] table [x=id, y=arg_ratio, col sep=comma] {data.csv};
				\addplot[style={mark=o}, skip coords between index={0}{30}] table [x=id, y=arg_ratio, col sep=comma] {data.csv};
			\end{axis}
		\end{tikzpicture}
		\label{fig:data_d}
	\end{subfigure}
	
	\caption{Collected data: Reflection score $\rue_{\argument_0}$, user stance $\effectiveness_{\argument_0}$, ratio of visited argument $\all_{v}^+:\all_{v}^-$ of both the experimental group (x) and the control group (o).}
	\label{fig:data}
\end{figure}

\subsection{Statistical Analysis}
\subsubsection{\rue~(H1):}

\begin{figure}[H]
	\begin{tikzpicture}
		\centering
		\begin{axis}[
			width  = 10.5cm,
			height  = 2.5cm,
			xbar,
			xmin=0.4,
			xmax=1.1,  
			axis lines*=left, 
			enlargelimits=0.0,
			enlarge y limits=0.5,
			bar width = 8pt,
			tickwidth = 0pt,
			legend style={at={(0.5,-0.75)},anchor=north, legend columns=4, font=\scriptsize},
			xtick distance=0.2,
			y label style={at={(axis description cs:-0.12,.5)},anchor=south, font={\scriptsize}},
			x label style={at={(axis description cs:0.5,-0.20)},anchor=south,font={\scriptsize}},
			xlabel={Value $\in [0,1]$},
			symbolic y coords={$RUE_{\Phi_0}$},
			ytick=data,
			tick label style={font={\scriptsize}},
			legend cell align=left,
			nodes near coords align={vertical},
			]
			\addplot+[style={fill=graph_green, text=black, draw=graph_green}, x filter/.expression={x==0 ? nan : x}, error bars/.cd, x dir=both, x explicit, error bar style=black] coordinates {
				(0.93,$RUE_{\Phi_0}$)  +- (0.027, 0.027)
			};
			\addplot+[style={fill=graph_red, text=black, draw=graph_red}, x filter/.expression={x==0 ? nan : x}, error bars/.cd, x dir=both, x explicit,error bar style=black] coordinates {
				(0.879,$RUE_{\Phi_0}$)  +- (0.0336, 0.0336)
			};
			\draw [arrows={Bar[right]-Bar[left]}, shorten <= 15pt, shorten >=5pt] 
			(axis cs:1,$RUE_{\Phi_0}$) ++(0, -10pt) -- node[midway, right, shift={(0,4pt)}]{*}  (axis cs:1,$RUE_{\Phi_0}$) ;
			\legend{w/ intervention, w/o intervention}
		\end{axis}
	\end{tikzpicture}
	\caption{User reflection score $\rue_{\argument_0}$.}
	\label{fig:results}
\end{figure}

Concerning the RUE scores (Fig.~\ref{fig:data_a}), the homogeneity of variances was falsified using the Levene's test($F=11.3$, $p=0.001$) and the assumption of a normal distribution using the Shapiro-Wilk test ($W=0.895$, $p < 0.001)$. Thus, we applied the \emph{Mann-Whitney-U test} to verify \ref{h:1} that the user reflection increased significantly in the experimental group (with intervention)($U = 273$, $n_1=30$, $n_2=28$, $p \leq 0.01$). In addition to that, we also checked the total amount of interventions. There were 262 interventions throughout the study (8.73 per user), 201 of which were accepted by the user, which is an acceptance rate of $76\%$. 


\subsubsection{Global Reflective Engagement (H2):}To test $H2$, 
We analyzed how many participants were more engaged with the opposing stance than with their own stance in both groups. Therefore, we compared the amount of pro and con arguments that the user heard to the user stance, e.g., if the user stance is negative ($ \effectiveness_{\argument_{0}} < 0.5$) and more pro than con arguments were heard ($\all_{v}^+ > \all_{v}^-$) this implicates a strong opposing engagement. We found that with intervention, nearly $80\%$ of participants were more engaged with the opposing stance, while only $53\%$ in the control group did so, which is a total increase of $51\%$. 

\section{Conclusion and Summary}\label{sec:summary}
Critical reflection raises the necessity to question the opposing view and relate it to one's own. To this end, we presented a novel intervention approach that considers the users' exploration behavior in dependence on their focus to increase reflective scrutinizing of arguments on controversial topics. We proposed an aggregated model to estimate the user's reflective engagement (RUE) based on the user's argument exploration focus and stance. We conducted an online user study to evaluate our approach, especially the intervention mechanism based on our model. The results for the experimental group showed a significant increase in reflective engagement than the control group ($U = 273, p \leq 0.01$). Furthermore, we found that the user's engagement with the opposing stance increased by about 51\%, indicating the practical potential of our proposed reflection model.

In future work, we will investigate how our model can be merged with implicit user feedback via social signals to develop a user-adaptive dialogue strategy to maximize RUE and user satisfaction. 

\section*{Acknowledgements}
This work has been funded by the Deut\-sche Forschungsgemeinschaft (DFG) within the project \enquote{BEA - Building Engaging Argumentation}, Grant Number 455911629, as part of the Priority Program \enquote{Robust Argumentation Machines (RATIO)} (SPP-1999).

\bibliography{main}
\end{document}